# Big Data Processing in Complex Hierarchical Network Systems

Olexandr Polishchuk, Dmytro Polishchuk, Maria Tyutyunnyk, Mykhailo Yadzhak

Department of Nonlinear Mathematical Analysis, Pidstryhach Institute for Applied Problems of Mechanics and Mathematics, National Academy of Sciences of Ukraine, Lviv, Ukraine

**Abstract.** This article covers the problem of processing of Big Data that describe process of complex networks and network systems operation. It also introduces the notion of hierarchical network systems combination into associations and conglomerates alongside with complex networks combination into multiplexes. The analysis is provided for methods of global network structures study depending on the purpose of the research. Also the main types of information flows in complex hierarchical network systems being the basic components of associations and conglomerates are covered. Approaches are proposed for creation of efficient computing environments, distributed computations organization and information processing methods parallelization at different levels of system hierarchy.

## 1. Introduction

Complex large scale technological systems (CLSTS) are used almost in all areas of human activity, e.g. in transportation (railway, road and aviation systems, transportation networks of large cities and regions of countries) [1], supply and logistics (systems for power, gas, petrol, heat and water supply, trade networks) [2], information and communication (Internet, TV, radio, post service, press, fixed and mobile telephony) [3], in economics (networks of state-owned and (or) private companies, their suppliers and final products distributors) [4], finance (banking and insurance networks, money transfer systems) [5], education, healthcare etc. Their state and operation quality impose large impact on citizens' quality of life, efficiency of economy and possibilities for its development, as well as government structures readiness to mitigate impacts of technological and natural disasters. Finally, they may be treated as the evidences of country development level in general [6, 7]. Failure of one of the elements of such system can often lead to operation breakdown or destabilization of the whole CLSTS. The example of this is cascading phenomenon [8]. Often the situations of the kind (e.g. accidents at nuclear or large chemical plants and other hazardous facilities, power lines, gas pipelines etc.) may lead to harsh consequences, such as environmental disasters, property loss and numerous human victims [9]. These circumstances determine the importance of continuous monitoring of technological systems operation, careful control of their behavior and timely response to emerging threats [10]. In this case, the major problem consists in the need to analyze large amounts of data that describe the state and behavior of CLSTS elements. This problem can be solved through creation of efficient computing environments, and usage of applied systems analysis and artificial intelligence methods [11, 12]. These methods implement effective algorithms for processing and analysis of information arriving from system components.

## 2. Complex Networks, Network Systems, Multiplexes, Associations and Conglomerates

During recent years, the theory of complex networks has been rapidly developing [13, 14]. We encounter network structures [15, 16] while studying micro- (e.g. quantum networks of fermions connected with bosons) and macroworld (gas networks in Universe, networks of black holes etc.). They occur in nature (e.g. protein and metabolism networks) and human society (e.g. Internet, language and citation networks). Complex technology systems (transportation, trade and power supply networks etc.) are not exceptions either. In general, an arbitrary network is defined as a statistical assembly, i.e. a set of networks with each network having certain probability of implementation, or as a set of all possible conditions of the given network. On the other hand, complex networks are graphs, i.e. the sets of nodes connected by some relations with nontrivial topological properties. When talking about real networks these properties determine network operation features [17].

Nodes of one network may be the nodes of many other networks at the same time. Thus, every town in the country can be a node for several transportation networks, as well as state and local administration networks, economy and financial network etc. Every person is also the node of many networks (family, professional, social etc.). Combination of several networks with non-empty intersection of nodes is called a multiplex [18]. Each network being the component of multiplex is called a layer. Examples above show that there are different types of interactions between the nodes existing on different layers of multiplex. These interactions may be of various nature or meaning and may have different material media.

Sometimes complex networks are called "the systems". In our opinion, network only reflects the structure of a system being its frame. There are flows that move along the network that make it a system. Real networks are created and exist with an aim to arrange the flows of certain type. These flows can be continuous (e.g. power resources), discrete (e.g. trains) or continuous-discrete (phone calls). The motion of flows in the network can be ordered (railway traffic), partially ordered (car traffic in large cities) or unordered (information flows in social networks). Networks with different types and levels of flow arrangement are generated different network systems [19]. Flow properties of certain network allow to divide multiplexes into network layers in a proper way.

CLSTS complies with any definition of a complex network and even network system only to some extent. The reason is that flows movement in majority of artificial technology networks need to be supported on organizational level. This function is performed by CLSTS control system which has hierarchical structure. Hierarchical network structures (HNS) are special because each subsystem of a certain hierarchy level consists of a set of subsystems which form subnetwork of lower hierarchy level network (see Fig. 1). Flow movement for which the network was created is performed at network of the lowest level. At the higher (control) levels, flows are represented by information, organizational and administrative decisions etc. HNS differs from common three dimensional tree structure by links between the nodes of each hierarchy level.

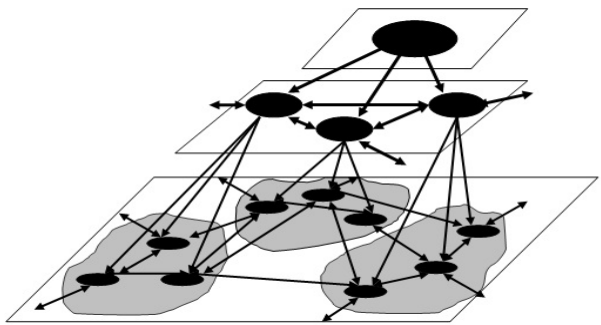

Fig. 1. Hierarchical network structure

There is an alternative to multiplex method of complex network systems combination. It consists in their joint engagement for solving important social or economical problems, e.g. industrial or natural disasters, pandemics, acts of terrorism etc. Solving these problems requires the interaction of many systems of different type and purpose: rescue and fire services, police, security agencies, military and medical units etc. The structures of the kind often arise in the industry, politics, social life etc. We call supersystems resulting from diverse complex hierarchical network systems (CHNS) interaction "conglomerates". Conglomerates often play more important role than multiplexes. Effective operation of conglomerates requires timely information exchange between their components. First of all, this means that interaction between CHNSs composing conglomerate has to be very tight. Relations between conglomerate components get less or more tighter depending on the purpose of their interaction and the extent of its implementation. Existence of purpose of creation and specific nature of relations are another aspects that make this structure different from multiplex.

The components of the conglomerate can comprise combination of several interacting systems of the same type. We call the structures of the kind "associations". Signs of uniformity determine whether some formation is a conglomerate or an association. If we talk about economy of the country in general, the combination of all transportation networks can be considered an association, since in this case the sign of uniformity is represented by the purpose of such system operation which lies in passengers and cargo transportation. At the same time, transportation system of the country is conglomerate of transportation systems of different types. Associations among components of such conglomerates are presented by transport operators organizations, transportation companies etc.

Sometimes the dilemma arises of whether some formation shall be considered a multiplex or a conglomerate. In general, it depends on purpose and extent of the study. Network of towns within the country is the basis for organization of flows of various types. From this point of view, it can be considered multiplex each layer of which provides a different type of flow movement (rail, road, air, sea, river), and each node of multiplex can support movement of up to all possible types of flows. However, the cooperation of transport systems of various types with an aim to transport passengers and cargo inside and outside the country can be considered conglomerate. Thus, the transportation system of the country can be considered multiplex, conglomerate or association within the larger conglomerate (for example, industrial) depending on the purpose of the study. Note that during the study of complex networks and multiplexes, structure properties have higher priority, and during the study of network systems, associations and conglomerates, the most important is the function implementing the purpose of their creation.

In general, the modern world is a huge dynamic multiplex-conglomerate structure with thousands of systems and billions of elements (nodes). Complex study of real multiplexes and conglomerates that constitute components of this structure requires complete and comprehensive understanding of their state and operation even in the case of their decomposition into network layers, associations and separate network systems. Such understanding can be achieved with the use of information about the history, current condition and forecast of systems behavior. The volume of this information, its diversity and problem of timely information flows content analysis lead to emergence of the phenomenon called Big Data.

## 3. Big Data and Information Overload

Big Data is the series of approaches and tools for processing huge volumes of structured and unstructured data. Properties of Big Data is defined by "three V": volume (large volume of data), variety (diversity of data), and velocity (processing speed and efficiency of results obtaining) [20]. The main goal of Big Data processing is to achieve the results that can be perceived by human. Before making decision, people try to get information that provides comprehensive characteristics of the problem being solved. However, as well as the lack of information, the excess of it can lead to achieving wrong results. Such situation is called information overload or the problem of analysis paralysis. It applies to CLSTS as well. In 1995, the study was carried out in the USA [21] with an aim to calculate the number of documents being processed by the employees of three large corporations. The result achieved was between 35 and 64 thousands. 20 years passed since that time. However, it is unlikely that information load on people working in those corporations, as well as other ones, has decreased since the volume and the number of information flows rise every year. It is important that the person responsible for operation of critical and dangerous facilities does not get lost in the flow of data describing the operation process of those facilities. Their task is to timely localize and solve the problem. To overcome information overload it is necessary to extract only the information essential for making correct and timely decision. Indeed, only small amount of data available to the person making decision satisfies this requirement. For example, the cracks in the railway rail have often led to accidents with numerous victims. The railway detector carriage used for cracks detection extracts the data with 1 mm step (in many countries, railroads are tens and hundreds of thousands kilometers long). As the result, multibillion arrays of numerical data [22] are obtained. At the same time, the only thing end user (maintenance unit) must know is the exact location of the crack. Many examples alike may be provided from power industry, economy and finance, medicine etc. The data regarding system elements behavior may continuously come in huge amounts from many sources and require real time processing. It is often necessary to store these data for long and short term CLSTS state and behavior forecasting. Many systems are sensitive to small changes the accumulation of which can be a threat to their normal operation. Analysis of stored information allows to identify the negative phenomena trends in advance and to prevent them beforehand.

The greatest problem of Big Data processing is induced by so-called unstructured information (text, photo, audio and video files). However, most of these data can be easily recognized and sorted by type, format, creation method etc. and sent to the relevant specialized tools for processing and analysis. Henceforth we will focus on processing of structured (in particular, numerical) data that usually make up the largest part of objective information about CLSTS operation process.

## 4. Information Flows in Complex Hierarchical Network Systems

We consider conglomerates as the combination of CLSTSs that interact to achieve common goals. Some CLSTSs can be grouped into associations according to uniformity. Every CLSTS is a complex hierarchical network system. At each level of the hierarchy, the edges ensure smooth motion of flows of certain type, whereas the nodes ensure their processing. Under the component of the system we understand its structural unit of any hierarchy level from element to subsystem of the highest decomposition level. Subsystems of the lowest decomposition level that consist of elements will be regarded to as the "basic subsystems" (BSS). The principles of the complex hierarchical network systems operation and methods of their behavior analysis are described in detail in [1].

In CHNS three main types of flows are distinguished (see Fig. 2):
1) ascending flows which come from controlled components to control components and may contain either processed primary data or aggregated data;
2) descending flows which come from control components to controlled components and contain information necessary for normal operation of controlled components, as well as decisions regarding their further actions;
3) network (horizontal) flows that come from some network components of certain hierarchy level to other components of the same hierarchy level.

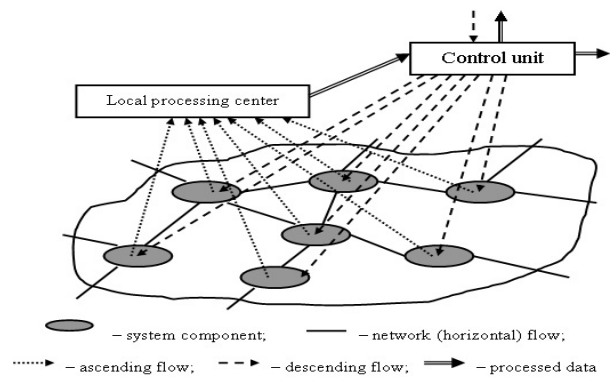

Fig. 2. Subsystem of CHNS

In general, ascending and descending flows implement cyclic (reverse) connection between control and controlled system components. These flows facilitate making correct and timely decisions regarding CLSTS operation process. Network flows provide self-organization of this process at every hierarchy level. These are the messages from one station regarding train delay to other stations located on the route of this train. Another example of such flows is professional information exchange in a team working on some project. In addition to the above mentioned, there are flows which run between the interacting systems of association or conglomerate. That is, we can add the intersystem flows to the listed intrasystem flows. One of the most important problems of CLSTS information support is synchronization of data flows between different CHNS hierarchy levels, within the networks of each hierarchy level and between the components of association or conglomerate.

We can also distinguish different levels of information processing. The first (the lowest) level consists in continuous extracting and real time processing of large volumes of numerical data that simultaneously arrive from many network elements (nodes) to local data processing and control centers. At this hierarchy level (level of system BSS) data analysis and timely response to detected local problems in the network elements (nodes, edges and flows) is carried out. The main purpose of information analysis at this level is to discover potentially failure (or "the weakest") BSS elements [23-26]. Failure of such elements often leads to the cascading phenomena (all catastrophes usually start from "minor failures" at system elements level). In such cases response time is major factor that allows to timely localize the problem and quickly overcome its consequences. Here it is expedient not only to determine current state but also to predict occurrence of potentially failure elements on the basis of previous evaluations history. In case such elements are missing, information processing center prepares the aggregated reports regarding condition of BSS elements and BSS in general for relevant control units. These reports are used as the basis for making objective and reasoned decision regarding further actions on this BSS. On this level, data on non-critical negative (which over time tends to become critical one) are accumulated for submission to upper control levels with an aim to solve existing problems and to prevent potential ones. In fact, at this level, objective information used to support decision-making by the control system units of the highest hierarchy levels shall be extracted and ordered.

Higher hierarchy levels receive the data aggregated in different ways [27]. Basic requirements for this data are as follows:
1) objectivity, i.e. they shall be based on reliable information only;
2) comprehensive description of the situation;
3) minimum sufficiency, i.e. the absence of duplicate and unimportant data;
4) simplicity and understandability, i.e. visualization methods shall be chosen that allow to quickly orient in the huge amount of conclusions obtained.

## 5. Hierarchical Network Computing Environments

Network technologies has been rapidly developing during recent 20 years [28, 29]. At the same time, network bandwidth continuously increases and the computer interaction latency decreases. Communication characteristics of many local (LAN) and wide area (WAN) networks have already exceeded the corresponding parameters of the first cluster computing systems. However, the process of network development continues. Due to the number of positive properties, they have become the basis for a new generation of computer systems – distributed computing environments (CE) [30, 31].

In each CLSTS, a distributed CE that can be called a kind of hierarchical network grid usually has been already created and it is developing. Structure of this hierarchical network computing environment (HNCE) in general reflects the structure of the system itself. Usually it is closed (in contrast to classical understanding of Grid [32]) information network where data exchange is regulated by quite strict rules. The hierarchy of the control system of CLSTS determines the hierarchy of user access rights to the system internal information. This can be explained by clear requirements of internal and external security, since the leakage or distortion of certain data can lead to unpredictable negative consequences [33, 34]. Indeed, it is hard to imagine situation where any user can change the schedule of trains on the railway. For many government agencies or large transnational corporations the scope of such HNCEs can significantly exceed the volume of traditional voluntary, scientific or commercial Grids. The structure and capacity of such CE must first of all provide the opportunity to work with Big Data, i.e. to store all the necessary information, provide high speed data access, support efficient algorithms for information processing, analysis and system behavior forecasting, ensure information security and protection from unauthorized access. Creating of HNCE is the first stage of the distribution and parallelization of Big Data processing in CLSTS.

This CE has parallel architecture and uses distributed memory, which makes it similar to computing clusters. However, the real essence of this environment is determined by the following main features:
- *distributedness*: the distance between computing nodes of this environment may be up to tens, hundreds or thousands kilometers, which leads to significant latency during their interaction;
- *heterogeneity*: nodes of environment may have different architecture and performance levels;
- *agility*: the architecture of the environment is able to change, new computing resources can be quickly added to it;
- *reliability*: any node can be extracted from the environment at any moment but this should not lead to delay or shutdown of the computing process;
- *large scope*: total number of processors (cores), general memory volume, number of users working simultaneously and applications running can be very large.

It is obvious that the distributedness of environment can lead to significant delays during remote nodes interaction. The only way to increase the efficiency of data processing in this case consists in maximum loading of nodes with computing tasks and minimizing the number of exchange operations between them.

Potential heterogenity of CE is predetermined by the fact that each node can process and store different volume of data and implement special algorithms of their processing. This means that computing node performance actually depends on the amount of data being processed, and its architecture – on the characteristics of highly parallel processing algorithms. Usually this is the cluster with required number of processors (or computing cores) and sufficient memory volume, including data warehouses. Clusters are known to be an affordable analogue of MPP (Massively Parallel Processing) systems and incorporate multiple SMP (Symmetric Multi-Processing) systems. Therefore, they can effectively implement highly parallel algorithms oriented at computers with common and distributed memory. In this case, we can use OpenMP (Open Multi-Processing) and MPI (Message Passing Interface) services of parallel programming to create parallel programs. If it is necessary to effectively carry out parallel algorithms that implement SIMD (Single Instruction Multiple Data) mode [35] or optimal performance parallel pipeline algorithms, the node of CE can have hybrid architecture. In this case, new computing units of graphics or quasisystolic processors can be connected to the cluster [36, 37].

It should be noted that the nodes have maximum commitment to CE, i.e. delegate it all their resources and work solely for it. Obviously, the structure of relations between nodes in environment is agile and can change in case separate nodes fail or new nodes are connected.

Reliability of CE means that in case one node fails or leaves the environment the whole computing operation of this node is assigned to another node of the same hierarchy level. The node that received additional tasks will probably require involving additional computing resources.

The scope of CE allows to integrate large number of independently operating computers, which means that it has powerful parallelism (and, therefore, productivity) resource. This is why such environment may be referred to as high performance CE.

In general, the structure of computing operation arrangement in such environment is based on independent processes. In this case, there is no common memory, common variables and other objects, and the interaction between the processes is carried out indirectly through message transfer. Thus, the CE described can be considered as an effective tool for the metacomputer computations [38] with the aim to solve complex preprocessing problems and further analysis of large data sets in CLSTS. In contrast to modern expensive supercomputers, it is not unique single device and its computing resources are used in an optimum manner.

## 6. Preprocessing of Large Data Sets

When working with large amounts of information we may encounter several important problems such as:
- providing high data transmission speed;
- creation of data warehouses and arrangement of high speed access to the data contained in them;
- real time processing;
- providing proper visualization of data processing results.

Solution of the problems above is impossible without the involvement of modern parallel computing systems. Such systems must be installed in every local processing center. It should be noted that first two problems can be solved by technology means. As to the fourth problem, approach to its solution can be based on the development of new principles of study outcomes visualization, which was partially performed in [22, 27].

Let us describe the approaches to solution of the third problem more detailed. First, we shall pay attention to data preprocessing. Information that comes from the elements of CLSTSs to local processing centers in the form of numerical data arrays may contain inaccurate or significantly distorted components. Therefore, before such arrays may be used for further analysis, they need to be preprocessed. We use for this methods and algorithms of digital filtering. In most cases, preprocessing must be carried out in real time. We suggest using quasisystolic method of computations arrangement for this purpose [39]. This method has become a basis for developing parallel pipeline algorithms of digital filtering in different dimensions with optimum performance and memory usage [40]. In this case, optimality was proved in the class of algorithms that are equivalent according to information graph within the accuracy of associativity and commutativity ratios performance. Proposed quasisystolic method has been extended to problems of cascade digital filtering in different dimensions [41] and problems of filtering that use adaptive smoothing procedure [42]. Optimum parallel pipeline filtering algorithms that were developed are oriented at implementation via special computer tools i.e. quasisystolic structures. These structures differ from systolic structures because they allow to perform data transmission from a single output point to multiple input points. Such quasisystolic structures can be considered as separate accelerating units in multipurpose parallel computing systems [43] used for solving filtering problems.

We have proposed and studied algorithms with limited parallelism [44] for solving digital filtering problems on systems with common or distributed memory, as well as on the systems with structural and procedural organization of computations. Note that in some cases, digital filtering problems in different dimensions may be solved with the help of neural network methods and algorithms [45] oriented on implementation on modern and prospective parallel computing systems.

## 7. Parallelization of Primary data Processing Procedures

Filtered data sets can be used to analyze the state and operation of system elements. While performing such analysis, the number of element characteristics, operation modes and evaluation criteria and parameters should be taken into account [46]. This leads to a substantial increase of computational expenditures. Therefore, to speed up the calculations it is necessary to develop approaches to their parallelization.

The procedure for local CLSTS element characteristic behavior evaluation was described in detail and analyzed in [47]. We have proposed parallel and sequential approach for carrying out this procedure in modern CE that use common memory. According to this approach, parallelization of some fragments is performed in the form of several autonomous branches, and the mode of parallelization of the other ones is close to full binary tree. To implement proposed approach we have developed algorithmic constructions, that take into account actual capabilities of available computing resources (the number of processors, computational cores, RAM volume etc.). We have also obtained results of parallel computing acceleration evaluation which confirm the efficiency of constructions developed.

## 8. Parallel Organization of Formation of Aggregated Conclusions

Aggregation allows to make generalised conclusions regarding the state and operation quality of both elements and separate sybsystems of a system on the basis of processed local data. To provide clarity and understanding of the aggregated behavior evaluation process for separate elements and CLSTS in general we have constructed corresponding analysis trees [48]. According to these trees we proposed and studied four types of algorithmic constructions for organization of parallel calculations. They represent two stages of parallelization. On the first stage, two constructions are described, as follows:

• parallel and sequential construction that consists of two consecutive fragments: the first one (the main, it assigns carrying out large amount of computations) is a set of certain number of parallel independent branches, the results of their carrying out is then used in the second fragment (with small amount of computations);

• parallel construction that is a set of some number of independent parallel branches performing approximately equal amount of calculations.

In the second stage of parallelization, more efficient algorithmic constructions are proposed. They are obtained from the two constructions described above in the following way: calculations in main fragment of each branch are carried out in the form of some number of independent parallel branches. These algorithmic constructions are oriented to implementation on parallel computing systems with common and distributed memory, in particular, on clusters. Cluster computing systems are affordable analogues of massively parallel systems as they are assembled from serial production components. This is why parallel algorithms for aggregated conclusions formation may be implemented with the help of properly configured clusters.

## 9. Future Research

We have used described in this article methods and approaches to optimize the structure and functioning of CE for such system as the regional railroad of the country. This CE is a prime example of HNCE. Now the work is done in two directions. The first direction is to extend the experience of optimization of regional railways CE to the HNCE of the railroad of the country as a whole. The second direction is to improve the interaction between regional rail and road cargo transport association, as the components of the transport conglomerate of region. Here the focus is on the synchronization of information flows between them, optimizing logistics operations and reduce delays in the movement and processing of local and transit cargo.